\documentclass{aa}
\usepackage{psfig}
\usepackage{rotating}

\begin{document}
\title{$CHANDRA$ Detection of 16 New X-Ray Sources}
\author{A. K\"upc\"u Yolda\c{s} \and \c{S}. Balman} 
\institute{Department of Physics, Middle East Technical University, 
In\"on\"u Bulvar{\i}, 06531 Ankara, TR}

\date{Received  / Accepted  }
\abstract{We have detected 18 sources over 6 $\sigma$ threshold within two 
regions 8$\arcmin$.3$\times$16$\arcmin$.9 and 8$\arcmin$.3$\times33\arcmin$.6 
in the vicinity of the point with $\alpha$=03$^h$31$^m$02$^s$.45 (J2000) 
and $\delta$=+43$\degr$47$\arcmin$58$\arcsec$.5 (J2000) using a 
$CHANDRA$ ACIS (S+I) observation. Two of the 
sources were detected before with $ROSAT$ HRI and one source could be 
closely identified with a star in the optical catalog, USNO A-2. We have 
also studied source spectra applying four spectral models to the data. 
Most of the sources can be classified as Cataclysmic Variable, Low Mass 
X-ray Binary or single star candidates due to their spectral 
characteristics and luminosities. We also searched for the extragalactic 
origin for these 18 sources. The source count rates vary between 
5.8$\times 10^{-4}$-4.7$\times 10^{-3}$ counts/s. Due to low count rates 
temporal characteristics of the sources can not be studied effectively.
}
\authorrunning{A. K\"upc\"u Yolda\c{s} \& \c{S}. Balman}
\titlerunning{CHANDRA Detection of New X-Ray Sources}
\maketitle

\section{Introduction}
The imaging capability and high sensitivity of the $Chandra$ $X-Ray$ 
$Observatory$ offers significant advantages for source detection. The 
good instrumental response for photon energies up to 8 keV allows 
detection of faint sources even if the source is subject to high 
absorption. The main goal of this search was to detect faint and Super 
Soft X-ray Sources (SSS) using a 95 ks observation obtained by the 
$CHANDRA$ Advanced CCD Imaging Spectrometer (ACIS) detector. The original 
target of the observation was a classical nova Persei 1901 (GK Per; 
principal investigator= \c{S}. Balman). The scientific aim of the 
proposal was deriving and studying the spectrum of the first classical 
nova shell resolved and detected in the X-ray wavelengths (Balman \& 
\"Ogelman \cite{bogel}). The $CHANDRA$ data reveals for the first time a 
classical nova evolving like a young, miniature supernova remnant. The 
scientific results of the $CHANDRA$ observation can be found in Balman 
(\cite{balman1}, \cite{balman2}). \\ 
As expected most of the sources we detected were faint sources, but we 
did not find any SSS above 5$\sigma$ confidence 
level. We analyzed the spectra of these new X-ray sources detected in two 
regions 8$\arcmin$.3$\times$16$\arcmin$.9 and 8$\arcmin$.3$\times33\arcmin$.6 
around the vicinity of the point with $\alpha$=03$^h$31$^m$02$^s$.45 
(J2000) and $\delta$=+43$\degr$47$\arcmin$58$\arcsec$.54 (J2000) within an 
energy range of 0.3-10 keV in order to identify their nature.\\ 
Details of the observation and the analysis methods are given in Section 
2. Section 3 discusses the detection properties of the sources and  
Section 4 is on the spectral analysis and discussion.\\
\section{The Data and Analysis}
The classical nova remnant of GK Per was observed with the $CHANDRA$ ACIS 
between 2000-02-10 and 2000-02-11.\\
ACIS  contains 10 planar, 1024 $\times$ 1024 pixel CCDs; four
arranged in a 2$\times$2 array (ACIS-I) used for imaging, and six arranged 
in a 1$\times$6 array (ACIS-S) used either for imaging or as a grating 
readout. Two CCDs are back-illuminated (BI) and eight are front-illuminated (FI). 
6 CCDs (2,3,5,6,7,8) have been used during our observation. GK Per is on 
CCD number 7. The level 2 data were processed using CIAO software (version 2.0) .
Bad pixels were removed and the latest suitable calibration files were 
used. In addition, we cleaned the data for the flaring effects of the 
ACIS-S3, and the exposure was reduced to 81 ks.\\
We ran the 
CELLDETECT\footnote{http://asc.harvard.edu/udocs/docs/swdocs/detect/html/} 
algorithm (or the `sliding cell' method), a source detection algorithm for 
X-ray data. It was developed for use with the Einstein Observatory images 
and also employed 
by the standard processing of ROSAT data (Harnden et al. \cite{harnden}). 
This method was tailored to optimize the detection of unresolved sources 
and had two variants, `local detect' and `map detect'. In the local 
detect that is used for our data, the background is estimated in a frame 
around the detect cell. At each point where a cell is placed, a 
signal-to-noise ratio of source counts to background counts is computed. 
If this ratio is above the detection threshold, a candidate source is 
recorded. The CELLDETECT method is good for faint point sources, outside 
crowded fields.\\ 
The CELLDETECT algorithm, run on the exposure corrected 0.3-10 keV images 
with exposure maps derived separately for five of the CCDs (2,3,5,6,7),  
yielded 20 sources over 6$\sigma$ detection threshold (6$\sigma$ 
above the background). Since CELLDETECT divides extended sources into 
multiple point sources and thus requires fine tuning of the parameters, we 
have chosen different values for the fixedcell parameter for different CCDs. 
The findpeaks parameter was set to `yes` so that the adjacent detections were 
recognized as a single source and the cell with the largest S/N appeared 
in the source list. We have omitted three of the sources and 
include one source detected with CELLDETECT algorithm over 12 $\sigma$ 
threshold at CCD 8. We have increased the threshold sigma of the 
detection algorithm up to 12 $\sigma$ for CCD 8 because of the excess 
number of bad pixels. Exposure maps for each of the CCDs were prepared 
using the single-chip exposure map thread of CIAO 2.0 and used to 
eliminate the false detections near the edges.
 
\section{Source Properties}
The results of our analysis with the CELLDETECT algorithm yield 18 sources 
over 6 $\sigma$ detection threshold. Table-1 displays all 18 sources 
including the source designations and count rates. The right ascension 
and declination errors determined from the CELLDETECT algorithm are 
between 0$\arcsec$.5 and 0$\arcsec$.2. The spatial resolution of 
$CHANDRA$ is 0$\arcsec$.49 pixel$^{-1}$. The source count rates vary in a 
range 0.5$\times$10$^{-3}$ - 5$\times$10$^{-3}$ counts/s.\\ 
After searching the HEASARC data  
archive\footnote{http://heasarc.gsfc.nasa.gov/docs/corp/data.html} 
including multi-wavelength catalogs and mission catalogs, two of the 
sources were found to be detected before with $ROSAT$ HRI and one source 
(Src 9) can be closely identified (with 0$\arcsec$.3 offset) with the 
source in the optical USNO A-2 Catalog (Monet et al. \cite{monet}).
The red and blue magnitude of the 
USNO source associated with Src 9 is m$_r$=14.7, m$_b$=16.2 . 
For offset radii up to 2$\arcsec$.5 we obtain 5 more possible candidate 
associations with the stars in USNO A-2 Catalog. However, we exclude 
these possibilities since the error in RA and DEC determined by 
CELLDETECT is lower than 2$\arcsec$.5. 
\begin{table*} 
\footnotesize
\caption{Identification Table Of The Sources (All coordinates are 
epoch=J2000)}
\begin{tabular}{llllll}
\hline
Src & Source$^\dagger$ & RA & DEC & {Counts/s $\times 10^{-3}$} & 
{$\sigma$$^\ddagger$} \\
\hline
1 & CXOAYSB J033025.9+434522.3 & 03 30 25.92 & +43 45 22.32 &
1.588$\pm$0.451 & 7 \\ \hline
2 & CXOAYSB J033037.0+434443.3 & 03 30 36.99 & +43 44 43.33 &
0.812$\pm$0.395 & 6 \\ \hline
3 & CXOAYSB J033017.9+435604.3 & 03 30 17.98 & +43 56 04.32 &
1.478$\pm$0.411 & 8 \\ \hline
4 & CXOAYSB J033121.0+434002.2  & 03 31 21.04 & +43 40 02.16 &
3.444$\pm$0.439 & 7 \\ \hline
5 & 1RXH J033136.5+434213 $^\star$ & 03 31 36.46 & +43 42 11.59 &
1.672$\pm$0.430 & 6 \\ \hline
6 & 1RXH J033102.3+434757 $^\star$ & 03 31 02.45 & +43 47 58.54 &
2.159$\pm$0.422 & 13 \\ \hline
7 & CXOAYSB J033056.1+434824.2 & 03 30 56.11 & +43 48 24.22 &
0.940$\pm$0.406 & 8 \\ \hline
8 & CXOAYSB J033128.6+435021.2 & 03 31 28.62 & +43 50 21.24 &
0.962$\pm$0.394 & 7 \\ \hline
9 & CXOAYSB J033122.2+435646.8 & 03 31 22.24 & +43 56 46.77 &
1.513$\pm$0.420 & 10 \\ \hline
10 & CXOAYSB J033117.7+435221.6 & 03 31 17.66 & +43 52 21.56 &
1.516$\pm$0.409 & 8 \\ \hline
11 & CXOAYSB J033108.3+435751.4 & 03 31 08.28 & +43 57 51.41 &
1.332$\pm$0.428 & 8 \\ \hline
12 & CXOAYSB J033045.0+435822.4 & 03 30 44.99 & +43 58 22.42 &
1.577$\pm$0.473 & 10 \\ \hline
13 & CXOAYSB J033131.6+435648.8 & 03 31 31.59 & +43 56 48.78 &
0.841$\pm$0.400 & 7 \\ \hline
14 & CXOAYSB J033105.2+435808.1 & 03 31 05.18 & +43 58 08.05 &
0.584$\pm$0.408 & 6 \\ \hline
15 & CXOAYSB J033057.1+435750.4 & 03 30 57.12 & +43 57 50.36 &
1.312$\pm$0.411 & 7 \\ \hline
16 & CXOAYSB J033118.3+435235.0 & 03 31 18.26 & +43 52 35.01 &
1.330$\pm$0.411 & 8 \\ \hline
17 & CXOAYSB J033113.3+435246.7 & 03 31 13.26 & +43 52 46.68 &
1.672$\pm$0.402 & 8 \\ \hline
18 & CXOAYSB J033106.0+440328.8 & 03 31 06.03 & +44 03 28.78 &
4.743$\pm$0.452 & 18 \\ \hline
\multicolumn{6}{l}{\ $^\star$ ROSAT Complete Results Archive Sources for
the HRI }\\
\multicolumn{6}{l}{\ $^\dagger$ The acronym CXOAYSB is registered in the
IAU Registry, see}\\ 
\multicolumn{6}{l}{\ http://cdsweb.u-strasbg.fr/cgi-bin/Dic?CXOAYSB} \\
\multicolumn{6}{l}{\ $^\ddagger$ Detection sigma above the background }
\end{tabular}
\end{table*}

\section{Spectral Analysis and Discussion}
The source spectra and insturment responses are generated using CIAO 
(version 2.0) and analyzed using XSPEC (version 11.0.1) applying four
models to the source data between 0.3 and 10 keV: blackbody, 
bremsstrahlung, powerlaw and VMEKAL with absorption. Spectra were grouped to 
have 5-10 counts per energy bin. The absorbed fluxes are found to be in a 
range 1$\times$10$^{-16}$ - 3$\times10^{-14}$ erg cm$^{-2}$ s$^{-1}$.\\
Table-2 shows the spectral parameters of the blackbody and
power law models and Table-3 shows the spectral parameters of the
bremsstrahlung and VMEKAL models. The errors on the spectral parameters
are at 2$\sigma$ confidence level. Figure 1 shows the fitted 
source spectra of all the 18 sources that has the best $\chi^2_\nu$s. The 
residuals in the figures are omitted since they were in a range of 
2$\sigma$ - {-2}$\sigma$, and did not show significant fluctuations.\\
Three of the sources (Src 4,12,16) are fitted using a power law model with 
photon indicies between 2.5 - 4 and have blackbody temperatures around 0.5 keV 
resembling the spectra of the Anomolous X-Ray Pulsars (AXPs) (Israel et 
al.\cite{israel}). At a distance of 10 kpc, 
the luminosities of these three sources are found to be around 10$^{32}$ 
ergs/s, and thus we exclude the possible AXP connection because the 
X-ray luminosities of AXPs are around 10$^{35}$ ergs/s.
Assuming a source distance of 10 kpc the luminosities of all the 18 
sources are calculated to be around 10$^{32}$ - 10$^{33}$ ergs/s. Such 
luminosities are consistent with galactic quiescent CV and quiescent LMXB origin.
Two of the sources (Src 11,13) show evidence for line emmision with
high absorbtion ( $N_H >$ 3$\times$ 10$^{21}$ cm$^{-2}$) and 
the best fit is the absorbed VMEKAL model. These sources are strong
Cataclysmic Variable and quiescent LMXB candidates (Warner \cite{warner};  
Verbunt et al. \cite{verbunt}; Guseinov et al. \cite{guseinov}). 
In addition, luminosities around 10$^{32}$ ergs/s could be attained by 
type O, B or giant stars as a consequence of shocks in the stellar winds or 
coronal emission (Cassinelli et al. \cite{cassinelli}; 
Schmitt et al. \cite{schmitt}). When we exclude the sources with 
bremsstrahlung temperatures above 1 keV (since almost all of O, B type or 
giant stars have temperatures below 1 keV), we are left 
with 3 candidates for galactic stars; Src 6, 9 and 12. 
In general, we reject an HMXB origin for our sources, since their 
luminosities are low compared with an HMXB where the luminosity is 
$\sim$10$^{36}$ ergs/s (Guseinov et al. \cite{guseinov}). The 
luminosities of the sources are around 10$^{37}$ 
ergs/s at 4 Mpc in agreement with the luminosities of the X-ray Binaries 
in other local galaxies, however we do not observe any host galaxy at 
those directions (Bauer et al. \cite{bauer}). As noted in Section 
1, none of these sources can be SSSs. The luminosities of the SSSs are 
around Eddington Luminosity (10$^{38}$ergs/s) and the blackbody 
temperatures are between 10-60 eV with almost no emmision above 1 keV 
(Kahabka \& van den Heuvel \cite{kahabka}). None of our 18 sources have 
such spectral characteristics.
For the case of Dim Thermal Neutron stars (DTNs) or cooling neutron stars,
the spectra of DTNs and cooling neutron stars are very soft and almost 
all of them are within 100 pc distance (Alpar \cite{alpar}; \"Ogelman 
\cite{ogelman}). A neutron star with a surface temperature around 10$^6$ 
K and a 10 km radius has an expected luminosity around 10$^{33}$ ergs/s. 
The observations on cooling neutron stars depend on the sensitivity of 
the X-ray telescopes. Among our 18 sources, the softest two (Src 6,11) 
have blackbody temperatures of 0.19 keV. This temperature is relatively 
high for a DTN or a cooling neutron star, however we do 
not exclude the possible DTN or cooling neutron star connection since the 
sensitivity of $CHANDRA$ allows us to observe fluxes around 10$^{-15}$ 
ergs cm$^{-2}$ s$^{-1}$ consistent with observation of a neutron star at 
10 kpc with a temperature $\sim$0.1 keV. 
Due to the relative hardness of the spectra of these 18 sources, we can 
also say that none of these sources are isolated hot white dwarf 
candidates since the temperatures of the hot white dwarfs are ultrasoft; 
a few eVs (Vennes \cite{vennes}; Finley et al. \cite{finley}).\\ 
  We searched for the extragalactic origins of 
all the 18 sources. The rest frame luminosities of AGNs and galaxies are
known to be between 10$^{39}$ and 10$^{45}$ ergs/s, and their spectra are
best fitted with power law models having photon indicies around 1.7 - 2
(Brandt et al. \cite{brandt}; Ishisaki et al. \cite{ishisaki}).
The rest frame luminosities of nearly all of the 18 sources are 
calculated to be around 10$^{42}$ - 10$^{44}$ ergs/s after fitting zmodels 
using XSPEC. Thus, we can not exclude the possibility of extragalactic 
origin for any of the sources. Two highly absorbed sources Src 3 and 5 
have rest frame luminosities around
10$^{42}$ ergs/s (redshift=0.206) and 10$^{43}$ ergs/s (redshift=0.500)
respectively. The photon indicies of these two sources are 1.61 and 1.89
respectively. Hence, we might categorize these two sources as strong AGN 
and galaxy candidates. Additionally, the N$_H$ values may be 
further evidence to exclude the extragalactic connection. The sources 
with N$_H$ higher than the value of the galactic N$_H$ in the direction 
of GK Per are more likely to be of extragalactic origin than the sources 
having the same order of N$_H$ with the galactic value. We may say that 
none of these sources are clusters since they are not extended and the 
rest frame luminosities of the clusters are usually higher than these values 
(Schindler \cite{schindler}).\\
In addition, we searched for temporal characteristics of the new sources. 
We performed 
power spectrum analysis on three of the sources with the highest count 
rates (Src 4, 6, 18). However we could not find any significant periods. 
The 3$\sigma$ - 4$\sigma$ detection threshold of power is around 40 in our 
data, and the power upper limit of the three sources is found to be 20. \\ 
Drawing conclusions about the classification of all the 18 
sources is difficult using the X-ray data at hand. We can not study their 
temporal characteristics due to low count-rates, and we do not have detailed 
spectra. We have planned further deep observations of this field in the 
optical wavelengths with the 1.5 m telescope of the National Observatory 
at Antalya, Turkey to ensure optical identification. Complementary 
observations in other wavelengths are necessary for proper classification 
of these 18 sources.

\begin{table*}
\vspace{2cm}
\begin{sideways}
\begin{tabular}{l|lllll|lllll}
\multicolumn{11}{l}{\ Table 2. Spectral Parameters Obtained Using Fits 
With Blackbody and Power Law Emmision Models}\\ \hline\hline
\multicolumn{1}{l|}{Source} &
\multicolumn{5}{c|}{\ Blackbody} &
\multicolumn{5}{c}{\ Powerlaw}\\
  & N$_H$ & kT & Norm $^1$ & Flux $^2$ & $\chi^2_\nu$ $^3$ & 
N$_H$ & PI & Norm $^1$ & Flux $^2$ & $\chi^2_\nu$ $^3$  \\
 & ($\times 10^{21}$ cm$^{-2}$) & (keV) & (counts/s/keV) & (ergs/cm$^2$ 
/s) & & ($\times 10^{21}$ cm$^{-2}$) & & (counts/s/keV) & (ergs/cm$^2$
/s) & \\
\hline
1 & 2.206 $^4$ & 0.30$^{+0.17}_{-0.08}$ & 1.49$^{+0.56}_{-0.40}$
& 1.04 & 1.66
& 2.370$^{+4.136}_{-2.286}$ & 2.31$^{+1.84}_{-0.93}$ & 49.27$^{<174.52}$ &
7.81 & 0.93\\
2 & 2.214 $^4$ & 0.82$^{+0.41}_{-0.29}$ & 1.14$^{+0.78}_{-0.49}$
& 8.56 & 1.78
& 0.023$^{<4.452}$ & 0.98$^{+1.09}_{-0.50}$ & 10.1$^{+14.9}_{-4.6}$ &
16.2 & 0.78 \\  
3 & 2.671$^{<8.432}$ & 1.03$^{+0.55}_{-0.32}$ & 3.28$^{+1.91}_{-1.17}$ &
21.8 & 0.71 &
9.675$^{+11.342}_{-5.948}$ & 1.65$^{+1.30}_{-0.92}$ & 75.12$^{<343.08}$ &
30.4 & 0.36 \\ 
4 & 0.586$^{<3.809}$ & 0.43$^{+0.22}_{-0.20}$ & 1.59$^{+1.57}_{-0.45}$
& 11.8 & 1.15 & 3.378$^{+3.572}_{-1.721}$ & 2.38$^{+0.55}_{-0.97}$ &
87.03$^{+177.04}_{-46.53}$ & 20.2 & 0.95 \\ 
5 & 0.392$^{<3.025}$ & 0.71$^{+0.23}_{-0.20}$ & 1.49$^{+0.76}_{-0.46}$
& 12.1 & 1.25 & 3.894$^{+5.184}_{-2.584}$ & 1.89$^{+1.19}_{-0.82}$ &   
50.38$^{+84.82}_{-26.38}$ & 19.5 & 1.30 \\ 
6 & 4.239$^{+4.596}_{-2.524}$ & 0.19$^{+0.06}_{-0.07}$ &
5.40$^{+54.44}_{-3.60}$ & 7.81 & 1.25 & 10.64$^{+5.94}_{-4.19}$ &
7.22$_{>5.81}$ & 866.49$^{+3619.08}_{-676.49}$ & 8.07 & 1.18 \\ 
7 & & & & & & 2.362$^{<9.798}$ & 1.31$^{+0.52}_{-1.04}$ &
13.06$^{<88.92}$ & 11.6 & 1.74 \\ 
8 & 2.296 $^4$ & 0.68$^{+0.28}_{-0.28}$ & 0.93$^{+0.51}_{-0.34}$
& 4.67 & 1.69 & 1.417$^{<6.813}$ & 1.50$^{+1.22}_{-0.73}$ &
16.13$^{<52.23}$ & 8.99 & 0.53 \\ 
9 & 0.974$^{<3.762}$ & 0.31$^{+0.17}_{-0.12}$ & 0.75$^{+1.49}_{-0.26}$
& 4.60 & 1.13 & 4.342$^{+4.188}_{-1.441}$ & 3.57$^{+2.60}_{-1.27}$ &
64.82$^{+177.78}_{-36.82}$ & 5.81 & 0.74 \\ 
10 & & & & & & 1.842$^{+3.371}_{-1.744}$ & 1.31$^{+1.02}_{-0.66}$ &   
18.0$^{+22.0}_{-8.0}$ & 16.4 & 0.68  \\ 
11 & 2.709$^{+10.572}_{-2.604}$ & 0.19$^{+0.15}_{-0.12}$ &
1.32$^{<2690.77}$ & 3.04 & 1.55 & 6.832$^{+7.771}_{-4.126}$ &
5.46$_{>2.39}$ & 106.1$^{<1017.5}$ & 3.21 & 1.61 \\  
12 & 2.382$^{+3.922}_{-2.164}$ & 0.29$^{+0.17}_{-0.12}$ &
1.25$^{+4.37}_{-0.55}$ & 0.75 & 1.08 & 6.471$^{+5.792}_{-3.197}$ &
4.18$^{+3.29}_{-1.61}$ & 139.9$^{+602.8}_{-93.9}$ & 1.93 & 0.80 \\
13 & & & & & & 11.26$^{+1.40}_{-1.34}$ & 9.99$_{>6.10}$ &
238.2$^{+176.8}_{-188.2}$ & 0.013 & 1.75 \\ 
14 & 2.071$^{<13.098}$ & 1.34$^{+1.56}_{-0.57}$ &
1.74$^{+5.12}_{-0.88}$ & 13.2 &
0.73 & 6.892$^{+16.827}_{-6.373}$ & 1.00$^{<2.64}$ &
14.11$^{+85.67}_{-10.61}$ & 18.4 & 0.56 \\ 
15 & 2.296 $^4$ & 0.76$^{+0.52}_{-0.41}$ &
1.09$^{+0.98}_{-0.54}$ & 8.02 &
1.89 & 2.956$^{+3.942}_{-2.438}$ & 1.46$^{+1.15}_{-0.51}$ &
19.8$^{+32.2}_{-9.8}$ & 13.9 & 0.79 \\  
16 & 0.826$^{<3.014}$ & 0.48$^{+0.17}_{0.16}$ & 0.85$^{+0.32}_{-0.26}$
& 6.25 & 0.95 & 5.680$^{+3.625}_{-2.866}$ & 3.27$^{+1.52}_{-0.88}$ &
77.5$^{+129.8}_{-45.5}$ & 6.88 & 1.07 \\ 
17 & & & & & & 1.642$^{+1.862}_{-1.406}$ & 1.28$^{+0.72}_{-0.50}$ &
20.3$^{+19.7}_{-7.8}$ & 16.2 & 0.81 \\ 
18 & & & & & & 2.795$^{+1.766}_{-1.112}$ & 2.10$^{+0.41}_{-0.36}$ &
141.44$^{+87.00}_{-50.42}$ & 45.9 & 1.58 \\ \hline
\multicolumn{11}{l}{\ $^1$ $\times 10^{-7}$, $^2$ $\times 10^{-15}$, $^3$ 
Fits that have {$\chi^2_\nu$$>$2} are excluded, $^4$ N$_H$ is fixed at 
the given value.}\\ 
\end{tabular}
\end{sideways}
\end{table*}

\begin{table*}
\vspace{2cm}
\begin{sideways}
\begin{tabular}{l|lllll|lllll}
\multicolumn{11}{l}{\ Table 3. Spectral Parameters Obtained Using Fits 
With Thermal Bremsstrahlung and VMEKAL Models}\\ 
\hline\hline
\multicolumn{1}{l|}{Source} &
\multicolumn{5}{c|}{\ Thermal Bremsstrahlung} &
\multicolumn{5}{c}{\ VMEKAL\ \ $^1$} \\
  & N$_H$ & kT & Norm $^2$ & Flux $^3$ & $\chi^2_\nu$ $^4$ & N$_H$ & kT &
Norm $^2$ & Flux $^3$ & $\chi^2_\nu$ $^4$ \\
  & ($\times 10^{21}$ cm$^{-2}$) & (keV) & (counts/s/keV) & (ergs/cm$^2$
/s) & & ($\times 10^{21}$ cm$^{-2}$) & (keV) & (counts/s/keV) & 
(ergs/cm$^2$/s) & \\ \hline
1 & 1.226$^{<3.843}$ & 2.92$^{+23.54}_{-1.81}$ &
50.85$^{+148.36}_{-30.65}$ & 7.40  & 0.96 & & & & \\ 
2 & 0.486$^{<4.698}$ & 200.0$_{>3.49}$ & 39.64$^{+20.17}_{-20.64}$ &
14.4 & 0.88 & 0.728$^{<5.008}$ & 94.95$^{+5.05}_{-91.17}$ &
79.28$^{+48.77}_{-28.80}$ & 13.1 & 0.91 \\ 
3 & 9.004$^{+7.438}_{-4.916}$ & 9.61$_{>2.43}$ & 84.62$^{<186.56}$ &
27.3 & 0.38 & 7.931$^{+9.588}_{-3.810}$ & 17.46$_{>2.56}$ &
234.71$^{+126.80}_{-78.86}$ & 33.2 & 0.44 \\ 
4 & 2.174$^{+2.464}_{1.304}$ & 2.76$^{+17.05}_{-1.54}$ &
80.77$^{+346.09}_{-30.77}$ & 18.6 & 1.00 & 9.131$^{+3.222}_{-3.615}$   
&
0.97$_{>0.46}$ & 217.33$^{+476.92}_{-64.71}$ & 9.04 & 1.84 \\ 
5 & 3.252$^{+3.693}_{-1.680}$ & 4.50$^{+118.34}_{-2.98}$ &
53.44$^{+59.32}_{-21.27}$ & 16.3 & 1.25 & 9.9$^{<17.4}$ &  
1.72$_{>0.97}$ & 205.7$^{<389.0}$ & 11.3 & 1.74 \\ 
6 & 5.689$^{+4.831}_{-2.232}$ & 0.34$^{+0.20}_{-0.18}$ &
2311.5$^{+218319.7}_{-1897.9}$ & 8.02 & 1.26 &
2.048$^{11.068}_{-1.941}$ &
1.06$^{+0.33}_{-0.83}$ & 60.11$^{+27419.46}_{-16.55}$ & 7.32 & 1.40 \\
7 & 2.184$^{+4.186}_{-1.926}$ & 46.03$_{>2.61}$ & 25.020$^{<51.267}$ &
11.4 & 1.76 & 3.034$^{+4.680}_{-2.562}$ & 5.72$_{>1.50}$ &
53.52$^{+37.32}_{-20.22}$ & 8.06 & 1.79 \\ 
8 & 1.142$^{<5.122}$ & 13.18$_{>1.99}$ & 23.88$^{+27.22}_{-10.29}$ &
9.01 & 0.59 & 1.292$^{<6.041}$ & 9.48$_{>2.40}$ &  
56.71$^{+41.09}_{-19.98}$ & 7.96 & 0.52 \\ 
9 & 2.412$^{+2.764}_{-1.243}$ & 0.99$^{+1.16}_{-0.36}$ &
75.27$^{+867.55}_{-23.27}$ & 5.33 & 0.86 & & & & & \\ 
10 & 1.682$^{+2.326}_{-1.113}$ & 52.78$_{>3.23}$ & 35.3$_{>21.0}$ & 
16.1 & 0.69 & & & & & \\ 
11 & 3.102$^{+4.436}_{-1.869}$ & 0.54$^{<1.65}$ &
153.46$^{<2.12}$$^\times$$^{10^7}$
& 3.32 & 1.57 & 0.926$^{+2.816}_{-0.770}$ &
1.02$^{+0.42}_{-0.86}$ & 19.8$^{+18.6}$ & 3.23 & 1.25 \\ 
12 & 4.118$^{+3.824}_{-2.055}$ & 0.71$^{+1.26}_{-0.43}$ &
197.8$^{+3873.5}_{-149.8}$ & 1.29 & 0.92 &
8.778$^{+5.212}_{-3.359}$ & 0.74$^{+0.33}_{-0.51}$ & 163.88$^{<3523.57}$ &
1.31 & 1.30 \\ 
13 & & & & & & 3.476$^{<8.587}$ & 0.30$^{+0.58}_{-0.16}$ &
49.50$^{<4756.15}$ & 0.016 & 1.35 \\ 
14 & 8.402$^{+10.524}_{-4.542}$ & 52.84$_{>0.0001}$ &
40.5$^{+45.7}_{-19.5}$ &
15.8 & 0.61 & 8.265$^{+11.180}_{-4.252}$ &
80.02$_{>2.80}$ & 120.38$^{+65.20}_{-49.60}$ & 16.4 & 0.60 \\ 
15 & 2.501$^{+3.021}_{-1.440}$ & 29.04$_{>2.79}$ & 31.5$_{>20.0}$ &
14.8 & 0.80 & & & & & \\ 
16 & 3.421$^{+2.600}_{-1.501}$ & 1.45$^{+1.70}_{-0.78}$ &
66.2$^{+168.4}_{-36.2}$ & 6.78 & 0.91 & & & & \\ 
17 & 1.60$^{+1.98}_{-1.01}$ & 46.55$_{>4.16}$ &
40.7$^{+29.3}_{-12.7}$ & 15.4 & 0.80 & & & & & \\ 
18 & 1.794$^{+1.373}_{-0.897}$ & 4.12$^{+5.08}_{-2.53}$ &
130.90$^{+63.50}_{-28.60}$ & 41.8 & 1.60 & & & & & \\ \hline
\multicolumn{11}{l}{\ $^1$ Solar abundances are assumed, $^2$ $\times 
10^{-7}$, $^3$ $\times 10^{-15}$, $^4$ Fits that have {$\chi^2_\nu$$>$2} 
are excluded.}\\  
\end{tabular}
\end{sideways}
\end{table*}

\begin{figure*}
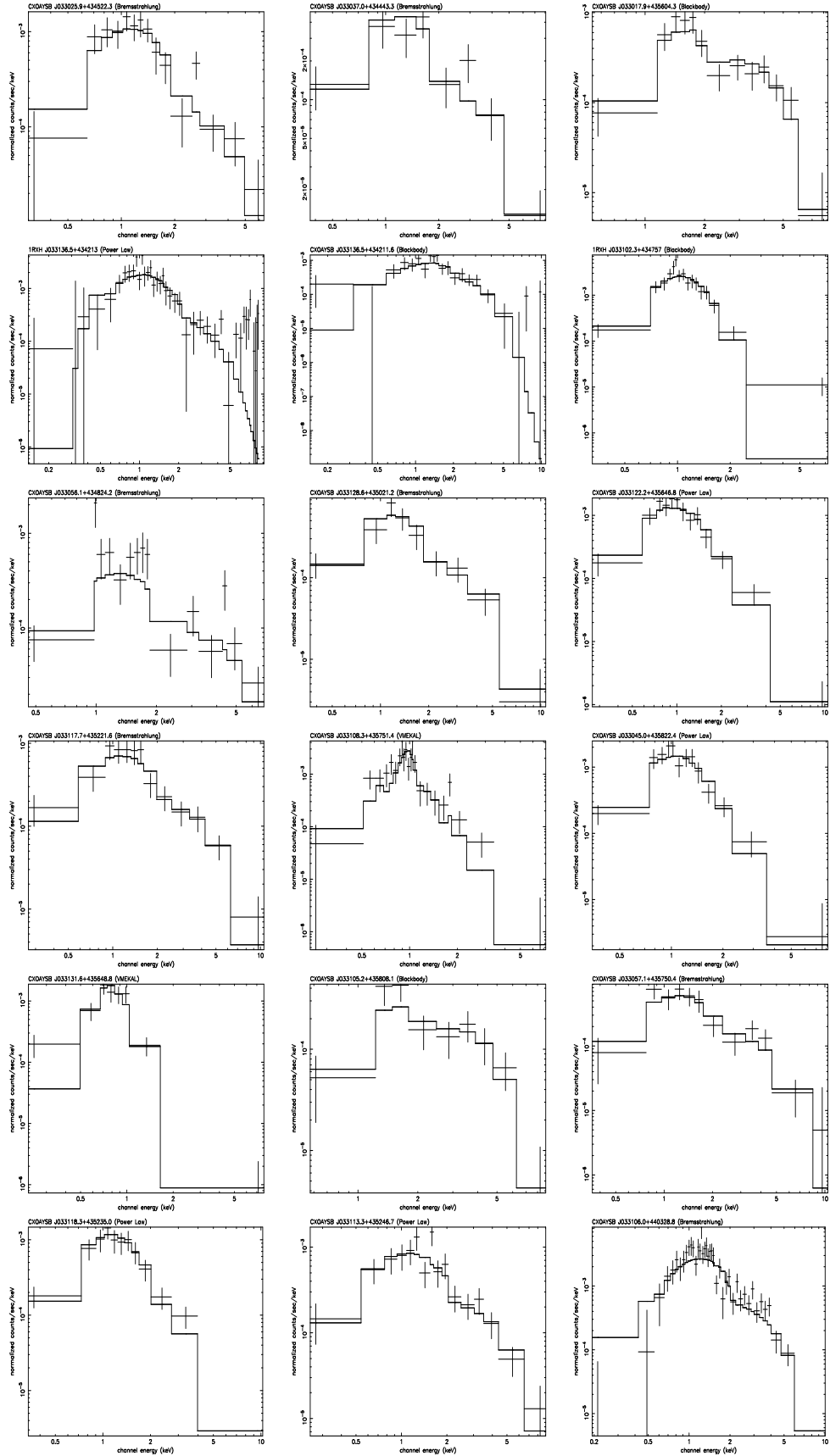

\footnotesize
\begin{center}
\begin{tabular}{ccc}
\psfig{file=h3378s1.ps,width=4cm,height=3.70cm,angle=-90} &
\psfig{file=h3378s2.ps,width=4cm,height=3.70cm,angle=-90} &
\psfig{file=h3378s3.ps,width=4cm,height=3.70cm,angle=-90} \\
\psfig{file=h3378s4.ps,width=4cm,height=3.70cm,angle=-90} &
\psfig{file=h3378s5.ps,width=4cm,height=3.70cm,angle=-90} &
\psfig{file=h3378s6.ps,width=4cm,height=3.70cm,angle=-90} \\
\psfig{file=h3378s7.ps,width=4cm,height=3.70cm,angle=-90} &
\psfig{file=h3378s8.ps,width=4cm,height=3.70cm,angle=-90} &
\psfig{file=h3378s9.ps,width=4cm,height=3.70cm,angle=-90} \\
\psfig{file=h3378s10.ps,width=4cm,height=3.70cm,angle=-90} &
\psfig{file=h3378s11.ps,width=4cm,height=3.70cm,angle=-90} &
\psfig{file=h3378s12.ps,width=4cm,height=3.70cm,angle=-90} \\
\psfig{file=h3378s13.ps,width=4cm,height=3.70cm,angle=-90} &
\psfig{file=h3378s14.ps,width=4cm,height=3.70cm,angle=-90} &
\psfig{file=h3378s15.ps,width=4cm,height=3.70cm,angle=-90} \\
\psfig{file=h3378s16.ps,width=4cm,height=3.70cm,angle=-90} &
\psfig{file=h3378s17.ps,width=4cm,height=3.70cm,angle=-90} &
\psfig{file=h3378s18.ps,width=4cm,height=3.70cm,angle=-90} \\
\end{tabular}
\caption{Fitted Spectra of the 18 Sources Using the Data from the 
$CHANDRA$ ACIS Detector Between 0.3 - 10 keV Energy Range}
\end{center}
\end{figure*}


\begin{thebibliography}{}
\bibitem[2001]{alpar}
  Alpar, M. A. 2001, ApJ, 554, 1245
\bibitem[2002]{balman2}
  Balman, \c{S}. 2002, in ASP Conf. Ser. , High Energy Universe at Sharp 
  Focus: Chandra Science, ed. S. Vrtilek, E. Schlegel, L. Kuhi, in press 
\bibitem[2001]{balman1}
  Balman, \c{S}. 2001, in ASP Conference Ser. 234, X-ray Astronomy 
  2000, ed. R. Giacconi, L. Stella, S. Serio, 269
\bibitem[1999]{bogel}
  Balman, \c{S}., \"Ogelman, H. B. 1999, ApJ, 518, L111
\bibitem[2001]{bauer}
  Bauer, F. E., Brandt, W. N., Sambruna, R. M., et al. 2001, AJ, 122, 182
\bibitem[2001]{brandt}
  Brandt, W. N., Hornschemeier, A. E., Alexander, D. M., et al. 2001, AJ, 
  122, 1
\bibitem[1981]{cassinelli}
  Cassinelli, J. P., Waldron, W. L., Sanders, W. T., et al. 1981, ApJ, 
  250, 677   
\bibitem[1997]{finley}
  Finley, D. S., Koester, D., \& Basri, G. 1997, ApJ, 488, 375
\bibitem[2000]{guseinov}
  Guseinov, O. H., Saygac, A. T., Allakhverdiev, A., et al. 2000, Astro. 
  Lett., 26, 725
\bibitem[1984]{harnden}
  Harnden, F. R., et al. 1984, SAO Report No. 393
\bibitem[2001]{ishisaki}
  Ishisaki, Y., Ueda, Y., Yamashita, A., et al. 2001, PASJ, 53, 445
\bibitem[1999]{israel}
  Israel, G. L., Covino, S., Stella, L., et al. 1999, ApJ, 518, L107
\bibitem[1997]{kahabka}
  Kahabka, P. and van den Heuvel, E. P. J. 1997, ARA\&A, 35, 69
\bibitem[1998]{monet}
  Monet, D., et al. 1998, The PMM USNO-A2.0 Catalog (Washington,
  D.C.: U.S. Naval Observatory)
\bibitem[1995]{ogelman}
  \"Ogelman, H. 1995, in The Lives of the Neutron Stars, ed. M. A. Alpar, 
  \"U. K{\i}z{\i}lo\u{g}lu, \& J. van Paradijs (Dordrecht: Kluwer), 101
\bibitem[1999]{schindler}
  Schindler, S. 1999, A\&A, 349, 435
\bibitem[1993]{schmitt}
  Schmitt, J. H. M. M., Zinnecker, H., Cruddace, R., Harnden, F. R., Jr. 
  1993, ApJ, 402, L13
\bibitem[1999]{vennes}
  Vennes, S. 1999, ApJ, 525, 995
\bibitem[1997]{verbunt}
  Verbunt , F., Bunk, W. H., Ritter, H., Pfeffermann, E. 1997, A\&A, 327, 602
\bibitem[1995]{warner}
  Warner, B. 1995, Cataclysmic Variable Stars (Cambridge:
  Cambridge University Press)
\end{thebibliography}
\end{document}